\documentstyle[prl,aps,multicol]{revtex}

\begin{document}

\begin{multicols}{2}

\noindent
{\bf BELL'S INEQUALITY:\hfill \break MORE TESTS ARE NEEDED}
\vskip .4cm

In the last two decades there have been numerous tests of unusual
correlations predicted by quantum mechanics between outcomes of
particular experiments in space-like separated regions. The peculiarity
of these correlations is that they are stronger than any correlations
explainable by a local theory. The quantum correlations,  as was proven by Bell
\cite{Bell}, break certain inequalities which have to be fulfilled  if the
results of every experiment are determined by some local hidden variable (LHV) theory.

We have been witnessed an outstanding progress in the tests of Bell's
inequalities, but a decisive experiment which would rule out any LHV
theory has not been performed yet. The deficiencies of the experiments
are described by {\it locality} and {\it detector efficiency}
loopholes. According to recent reports in Nature by Aspect \cite{Asp99} and
Grangier \cite{Gran} the two loopholes were closed.

 I want to express an opinion, that 
 the really important task in this field has not been  tackled 
yet and that the leading experiments claiming to close the loopholes, Weihs et 
al. \cite{nolo} and Rowe et al. \cite{delo}, although clearly making a very
significant progress, have conceptual drawbacks. These drawbacks have to be removed 
before claims like ``the last loophole closes'' can be made.

Today there is a firm consensus that there is no real question what
will be the outcome of this type of experiments: the predictions of quantum
theory or results conforming with the Bell inequalities. Predictions of
quantum mechanics were verified in so many experiments and with such
unprecedented precision that, in spite
of the very peculiar and nonintuitive features that Bell-inequality experiments demonstrate, only a minute minority of physicists
 believe that  quantum mechanics might fail in this type of experiments.
However, the fact that we are pretty sure about the final result of
these experiments does not mean that we should not perform them. One
goal of such experiments is to change our intuition which developed
from observing classical phenomena. But more importantly, these
experiments should lead to the stage in which we will be able to {\it
  use} these unusual correlations. 

Conceptually, the most simple, surprising, and convincing out of the
Bell-type experiments is the Mermin's version of the
Greenberger-Horn-Zeilinger (GHZ) setup \cite{Mer}.  I find that it can
be best explained as a game \cite{GHZV}. The team of three players
is allowed to make any preparations before the players are taken to
three remote locations. Then, at a certain time, each
player is asked one of two possible questions: ``What is $X$?''  or
``What is $Y$?'' to which they must quickly give one of the answers:
``$1$'' or ``$-1$''.  According to the rules of the game, either all
players are asked the $X$ question, or only one player is asked the
$X$ question and the other two are asked the $Y$ question. The team
wins if the product of their three answers is $-1$ in the case of
three $X$ questions and is $1$ in the case of one $X$ and two $Y$
questions.  It is a simple exercise to prove that if the answer of
each player is determined by some LHV theory, then the best strategy
of the team will lead to  75\% probability to win. However, a
quantum team equipped with ideal devices can win with certainty. So,
in my opinion, the most convincing test of the Bell inequality is
actually constructing such devices an seeing  that, indeed, the team wins the
game with probability significantly larger than 75\%. 

 Such an
experiment, if successful, will definitely close the
detection efficiency loophole. If it is also arranged that the party
 asking the questions chooses them ``randomly'' (more about
randomness below), then it will also close the locality loophole. Note,
that an experiment which simultaneously closes both the detection
efficiency and the locality loophole will not necessarily be suitable for
winning games of the type I described here. The essential property of
the experiment which allows winning games is that when one player
observes the result of his measurement, he has to be sure that other
players observe corresponding results too. We might imagine an
experiment with 100\% efficient detectors and ultrafast switches, but
with the source that  do not always send the GHZ triplets (like in
the current GHZ experiment \cite{GHZe}); such devices cannot help in
winning the game.

The technology for ``winning games'' experiments is not quite here
yet, so we should still work on ``closing the loopholes'' experiments.
The locality loophole means that the decision of the choice of which
local measurement to perform is made long enough before the time of 
measurements in the other sites such that the information about this
choice can be known there. Then, there is no difficulty in
constructing a LHV theory which reproduces the quantum correlations.
In the experiment which claimed to close the locality loophole, the
choices of the local measurements were determined shortly before the
measurement by fast quantum experiments at each site. The outcomes of
these experiments are ``genuinely random'' according to the standard
quantum theory.  However, if we want to rule out LHV theories, we
should consider the situation in the framework of such theories. The
outcomes of the  quantum measurements which determine the choice of
the local measurements  are also governed by some LHVs in
each site. There is enough time for information about these LHVs to reach other
sites before the measurements there took place and, therefore, a
consistent LHV theory which reproduces the results of Weihs et al.
experiment can be constructed. The locality loophole is not closed.  The experiment, nevertheless, is a
significant step forward  because its results can be explained only by
a higher level LHV  theory in which hidden
variable specifying the behavior of one system are influenced by
hidden variables of other systems.

A frequently discussed experiment in which the distance between the
sites is so large that a person at each site will have enough time to
exercise his ``free will'' to choose between the measurements will be
very convincing, but conceptually, not much better: we cannot rule out
the existence of LHVs which are responsible for our seemingly ``free''
decisions. A better experiment for closing the locality loophole
(which does not seem to be impossible with today's technology) is to
arrange that the choice of local measurements will be determined by
photons arriving from distant galaxies from opposite directions.  Then,
the only explanation will be a ``conspiracy'' LHV theory on the
intergalactic scale.

Finally, I will  discuss the latest experiment by  Rowe et al. \cite{delo}
who claimed to close the detection efficiency loophole. In this
experiment the quantum correlations were observed between results of
measurements performed on two ions few micrometers apart.
The detection
efficiency was very high. It was admitted that the locality
loophole was not closed in this experiment. But the situation about
locality was  worse than that. Again, the locality loophole means
that in principle, the information about the choice of local measurement can
reach the other sites before the time of measurements there. However,
in the Rowe et al. experiment not only the choice, 
but also the {\it result} of  the local measurement could reach the other
site before the measurement there was completed.
 The reading of the results was based on
observing numerous photons emitted by the ions. This process takes
time which is a few orders magnitude larger than the time it takes for
the light to go from one ion to the other. Thus, we can construct a
very simple hidden variable theory which arranges quantum correlations
by ``communicating'' between the ions during the process  of
measurement.

The purpose of closing the detection efficiency loophole was to rule
out set of LHV theories in which the particle carries the instructions
of the type: if the measuring device has particular parameters, choose
``up'', for some other parameters choose ``down'' and yet for some
other parameter choose ``not to be detected''. Such hidden variables
cannot explain the correlations of Rowe et al. experiment and this is
an important achievement.  However, the task of performing an
experiment closing the detection efficiency loophole without opening
new loopholes is still open.

 It is a pleasure to thank  Jonathan Jones and Lucien Hardy for helpful discussions.
 This research was supported in part by grant 471/98 of the Basic
 Research Foundation (administrated by the Israel Academy of Sciences
 and Humanities) and the EPSRC grant  GR/N33058.

\vskip .5cm
\noindent
{\bf Lev Vaidman}

\footnotesize
\vskip .3cm
\noindent 
{\it  Centre for Quantum Computation \hfill \break
 Department of Physics, University of Oxford\hfill \break
Clarendon Laboratory, Parks Road, Oxford OX1 3PU, England\hfill \break
and \hfill \break
 School of Physics and Astronomy\hfill \break
Raymond and Beverly Sackler Faculty of Exact Sciences\hfill\break
Tel-Aviv University, Tel-Aviv 69978, Israel}
\vskip .1cm
\noindent
e-mail: vaidman@post.tau.ac.il

\end{multicols}

\end{document}